\documentclass{aa}  

\usepackage{graphicx}
\usepackage{txfonts}
\usepackage{lipsum}
\usepackage{subcaption}         
\usepackage{lscape}             
\usepackage{placeins}           
                                
\def\nct#1{\nocite{#1}}

\begin{document}

   \title{Nonbirefringent model of orthogonal polarization modes\\ in radio
pulsars}

   \subtitle{New view on S swing and mode structure in pulsar beam}

%
%
%

   \author{J. Dyks\corrauth{jinx@ncac.torun.pl}        
        \and A. Frankowski\email{torke@poczta.onet.pl}
        }

   \institute{Nicolaus Copernicus Astronomical Center, Polish Academy of
Sciences, Bartycka 18, 00-716 Warsaw, Poland
   }

   \date{Received Jul 9, 2026}

 
  \abstract
{Two orthogonal polarization modes observed in radio pulsar signals have long been attributed to proper modes of
wave oscillation in strongly magnetized plasma. Yet it has been 
shown recently that they show up readily for extended emission regions that
produce incoherently-superposed polarization signal.}
{In this paper we present a two-dimensional polarization model based on
incoherent  superposition of radio signals. }
{The model involves a single proper mode, say the O mode, but leads to the
appearance of two orthogonal polarization tracks 
and naturally produces the triple form of
polarization mode segregation in averaged profiles (central mode flanked
on boths sides
by another mode),
as well as the displacement of modes in latitude,
previously inferred from beam mapping.}
{In the case
of conal emission regions, the modelled polarization tends to mimic general
polarization properties of the rotating vector model (RVM).
However, the reason for this is the symmetry
of the emission region - not the usual projection of dipolar magnetic
azimuths. Thus the emerging
RVM parameters reveal geometry of the emission region, not of the
dipolar magnetic field.}
{The results strongly support the vital role of nonbirefringent modal effects in radio pulsar profiles. 
Two proper modes may not be needed to explain observations of two orthogonal polarization
tracks.}

   \keywords{pulsars: general --
                polarization --
                radiation mechanisms: non-thermal
               }

   \maketitle
\nolinenumbers

\section{Introduction}

The anisotropic medium of pulsar magnetospheric plasma allows for
propagation of orthogonal polarization modes (OPMs) that are considered as a paramount
example of birefringence in astrophysics (p.~44 in Tinbergen 1996). 
\nct{t05} 
These proper modes were investigated in a number of detailed studies
based on first principles or electromagnetism (Cheng \& Ruderman 1979;
Arons \& Barnard 1986; Melrose \& Stoneham 1977; Rafat et al.~2019; Verdon
\& Melrose 2008). \nct{cr79, ab86, ms77,
rmm2019, vm2008}
The propagation of
radio waves through the magnetosphere was investigated, with mode coupling 
(Lyubarskii \& Petrova 1998; Petrova 2006; Wang \& Li 2009; Hakobyan et
al.~2017) \nct{lp98, 
p2006, wl2009, hbp17}
recognized as a likely process responsible for distortions
of polarization angle (PA) and zones of strong 
circular polarization V (Weisberg \& Taylor 2002; Edwards \& Stappers 2004; Mitra et al.~2016; Dyks
et al.~2021; Cao et al.~2025). \nct{wt02, es04, mra2016, dwi21, cjd2025}
Plasma density rarefication and
near-tangential propagation were recognized as possible cause of the
coupling 
(Andrianov \& Beskin 2010; Wang et al.~2010). \nct{ab2010, wlh10} 
Observationally, polarization modes in pulsar profiles 
are often overlapping in pulse longitude, and only partialy separated
(Stinebring et al.~1984; Hankins \& Rankin
2010; Mitra et al.~2015; Noutsos et al.~2015;  Johnston
et al.~2024). 
\nct{scr84, hr10, mar2015, nsk15, jmk2024}
They often form a triple O-X-O pattern with the central mode surrounded by another mode. 
This has
been interpreted through refraction or scattering of the O-mode (Barnard \&
Arons 1986; Lyubarski \& 
Petrova 1998). \nct{ba86, lp1998}


On the other hand the observed complexity of polarization has been
heuristically interpreted as a mixed state, 
ie. a superposition of the proper modes. The superposition was either
incoherent (McKinnon \& Stinebring 1998; Melrose et al.~2006, van Straten
\& Tiburzi 2017) \nct{ms98, mmk2006, vst17}
or coherent (Edwards \& Stappers 2004; 
Dyks 2019; Jones 2016) \nct{es04, d2019, jon2016}
as well as involved both these possibilities (Oswald et al.~2023).
\nct{okj2023} 

Deep understanding of pulsar polarization is nevertheless limited. 
The unclear area involves basic phenomena such as 
strong deviations of PA curves from the S swing of the rotating vector model
(RVM, Radhakrishnan \& Cooke 1969; Komesaroff 1970), \nct{rc69, k70} the weak     
relation between the profile intensity $I$ and the polarization angle, between the $I$
profile and a dominant OPM, the profiles of
polarized fraction ($L/I$ or $I_{\rm pol}/I$), the location of zones of high circular
polarization. 
Even more enigmatic are examples of complex behaviour such as fast PA rotations
(Cao et al.~2025, Manchester et al.~1975), \nct{cjd2025, mth1975} PA arcs or torii
(Primak et al.~2022). \nct{pts2022}
The question of OPMs'role is either
explicit or involved/underlying many of these effects. 

It has been found recently (Dyks et al.~2026, hereafter DSF26)
\nct{dsf2026}
that
incoherent superposition of polarized signal tends to produce apparent OPMs even if
only one proper mode is emitted. The OPMs with orthogonal PA jumps can thus
appear from simple geometric effects, such as the signal averaging over symmetric emission region. 
While DSF26 focused on one-dimensional emission region (thin stream), early
two-dimensional results (fig.~6 therein) revealed similar nonbirefringent
OPMs persisting for regions with the wide wedge or conal geometry.
In this paper we present a more general model of the nonbirefringent OPMs. 
The model is based on identical geometric effect, 
however, in comparison to DSF26 \nct{dsf2026} the present approach differs in the
following: 1) the model explicitly
includes two dimensions of an extended emitter, 2) it is largely detached from
the sky-projected direction of magnetic field line planes (or planes of
fixed magnetic azimuth), though it can imply RVM-shaped PA curves, but of different origin, 
and 3) it is capable of involving OPMs in nonsymmetric emitters. 
After presenting the model in Section \ref{model}, we present results that
surprisingly exhibit major features of known radio pulsar polarization
(Section \ref{results}), then summary follows in Section \ref{discu}.

\section{The model}
\label{model}

The model assumes that radio waves are locally emitted into a beamlet of
limited angular width $\rho$. In a standard case that would be $\rho\approx1/\gamma$ where
$\gamma$ is the Lorentz factor of electrons emitting at observed frequency
$\nu$. However, we allow for a larger beamlet size (even profile-wide) 
given that the microphysical radiation pattern may possibly be considerably rescaled by
the mechanism of `Doppler magnification' as described for the bifurcated
emission component of J1012+5307 (Dyks 2023). \nct{d2023} The angle $\rho$ is then 
considered just as a maximum angular range from which radiation can reach
the line of sight. 
Specifically, for the beamlet shape our calculations use:
\begin{equation}
I_{\rm blt}=\exp\left(-0.5\kappa^2/\gamma^{-2}\right)
\end{equation}
where $\kappa$ is the distance of the beamlet axis from the line of
sight and $\gamma$ is a free scale-governing parameter.

\begin{figure}[ht!]
   \centering
   \includegraphics[width=\hsize]{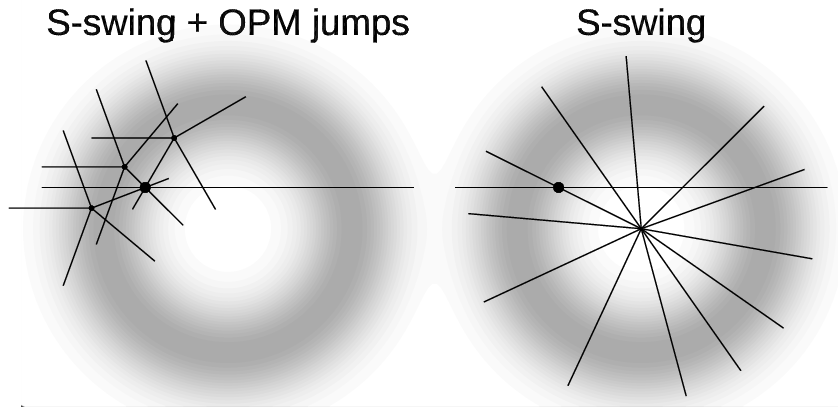}
      \caption{Comparison of our method of PA calculation (left) with the
RVM (right) for a hollow cone emission region. In our model polarization directions (as contributed by the
three exemplary emitting points) are radial around the line
of sight (bullet). The local average of such PAs produces the
S-swing along with jumps between two apparent OPMs. No OPMs show up in the RVM.
}
         \label{compar}
   \end{figure}

The model involves only one proper mode: the one polarized in the plane that
contains the wave vector $\vec k$ (assumed to be parallel to the line of
sight $\vec n$) and the local magnetic field $\vec B$ at the emission point. 
The wave vector $\vec k/|\vec k|=\vec n$ is orthogonal to the sky plane (and
can be imagined anchored at the emission
point) so the PA corresponds to the position angle of the sky-projected $\vec B$ at the emission point 
(the projection made along $\vec k$). 
In another (though equivalent) way the PA can be defined as follows:
A laterally-extended emission region can be projected on the sky along a
 local $\vec B$ field (ie.~the mapping to infinity is along the local $\vec
B$). The line of sight is a point on such a map. 
The angle of polarization $\psi$ is then determined by a line that connects the 
line of sight with such a sky-projected emission point. 
Our assumed polarization direction is contrasted with that of RVM in
Fig.~\ref{compar}. The relative orientation of $\vec k$ and $\vec B$
corresponds to locations close to the emission point, ie.~the possible ordering of polarization direction within 
the propagating beamlet, as caused by the increasing misalignment of local $\vec B$  
(adiabatic walking, see fig.~1a in Cheng \& Ruderman 1979) is neglected. 

The shape of an emission region (the region's local emissivity) is defined as
a function $I_{\rm er}$ defined on a two-dimensional net of
points $(x_i, y_j)$ within a rectangle. This may be a set of bright patches
(eg.~Gaussians), or a filled or hollow cone with Gaussian walls, etc.    
The pulse profile $I(\phi)$ forms when the region is cut horizontally at a
viewing parameter $y_{\rm obs}$ by the sightline.
At each pulse longitude $\phi_k$, the contribution from a given emission
point, expressed as Stokes parameters, is:
\begin{eqnarray}
I(x_i,y_j,\phi_k) & = & I_{\rm er}I_{\rm blt}\\
Q(x_i,y_j,\phi_k) & = & I_{\rm er}I_{\rm blt}\cos{2\psi}\\
U(x_i,y_j,\phi_k) & = & I_{\rm er}I_{\rm blt}\sin{2\psi}.
\end{eqnarray}
ie.~the beamlet emission is weighted by the regions' emissivity. 
The code runs over all points of the region to sum up all contributions:
\begin{eqnarray}
I & = & \sum_{i,j}I(x_i, y_j, \phi_k)\\ 
Q & = & \sum_{i,j}Q(x_i,y_j, \phi_k)\\
U & = & \sum_{i,j}U(x_i,y_j, \phi_k).
\label{sums}
\end{eqnarray}

Since the sizes 
involved are freely scalable (particularly beamlet size with
respect to the emitter's scale), the longitude $\phi$ is equivalent to the 
dimensionless horizontal coordinate $x$. The scale of emission region is 
assumed small so that spherical trigonometry is ignored and results are
valid for most inclinations of emission region with respect to the rotation
axis, except from very small ones. 
In the following we discuss a laterally-extended continuous emitter which is appropriate for 
averaged pulse profiles.

\subsection{Orthogonal modes from PA superposition in the vicinity of sightline}

   \begin{figure}[ht!]
   \centering
   \includegraphics[width=\hsize]{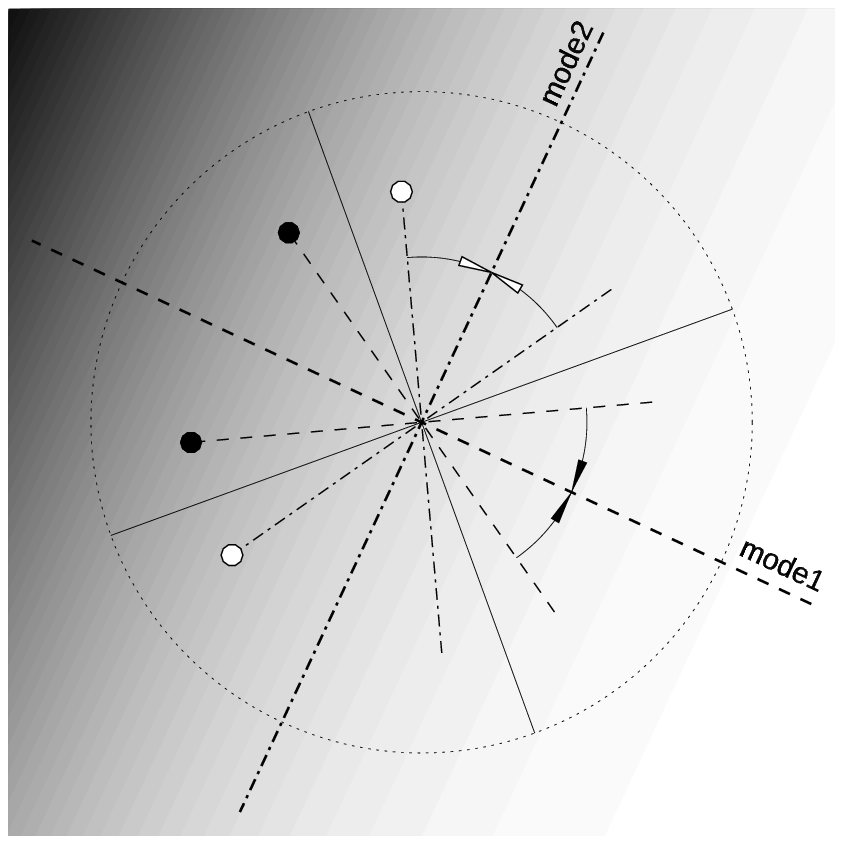}
      \caption{The mechanism of nonbirefringent OPMs. Grey scale shows the local emissivity projected
on the sky with dashed line marked `mode1' along the steepest gradient. 
The line of sight is a point at the center. Arbitrary emission point (bullets)
contributes radiation polarized radially, ie.~along a line connecting the point with the center. 
A pair of equally bright points (either black or white bullets)
gives a net PA in the nearest middle between the crossed polarizations of
these points, ie.~either along mode1 or
mode2, as shown with black or white arrows (also see
eqs.~\ref{resul1} and \ref{resul2}). Solid lines separate the 
quadrants of orthogonality, ie.~regions contributing to either mode; 
dotted circle limits the area in which radiation is detectable by 
the line of sight.}
         \label{lup}
   \end{figure}

Because of the limited beamlet range, the mechanism of nonbirefringent OPMs can be understood by considering
signal contributions from the vicinity of the line of sight. 
Fig.~\ref{lup} shows in grey a two dimensional emission region with the line of sight located at the
center. Dotted circle marks the furthest distance of detectable radiation (beamlet size).   
The dashed line of Mode 1 shows the steepest gradient of the
region's emissivity. Since for many regions the emissivity near a point can
be locally approximated by a
tilted plane, the gradient defines a plane of symmetry for the emissivity pattern
near the line of sight. Consider two equally-bright emission points (black bullets) located symmetrically
on each side of the thick dashed symmetry line. Because of the symmetry, the radiative contributions 
from the two black bullets average to the PA of Mode 1, as marked with black
arrows. This holds as long as the points fall within the thin solid
separatrices that cut the dashed symmetry line at $\pm45^\circ$. The
separatrices define four quadrants of orthogonality, one pair of quadrants aligned with
the steepest gradient, the other one orthogonal to the gradient. 
 Whenever the emission points are located further from the gradient line than the thin solid
separatrices (as the white bullets are), their net radiative input produces the
orthogonal PA of Mode 2, as marked by the white arrows. Each pair of symmetric points 
within the gradient-aligned quadrants contributes the PA of Mode 1 to the
signal, whereas each pair from beyond the quadrants contributes the PA of
Mode 2. There is a base PA value that corresponds to the gradient direction,
whereas the question of which mode is observed depends on how much net polarized flux is 
gathered by Stokes-wise integration within each pair of quadrants (pairs of
points located closer to the separatrices are subject to stronger
depolarization). 
 In a more general case (not shown in the figure), for a moving line of sight, 
the direction of the gradient rotates to
keep pointing to the brightest side, whereas the ratio of net polarized
flux integrated within the quadrant pairs can change and get inversed. This corresponds to the OPM jump.

As a simple geometric effect, the nonbirefringent OPMs can be easily derived
mathematically.
Let the PA of the gradient line in Fig.~\ref{lup} is denoted
as $\chi$ (dashed line marked `mode1'), and each black bullet (say with index 1
or 2) contributes radiation of the same intensity $I_b$ that is polarized at 
a positive angle $\Delta<45^\circ$ with respect to
the gradient line, so we have $U=U_1+U_2$ and $Q=Q_1+Q_2$, where:
\begin{eqnarray}
U_1 & = & I_b \sin(2\chi-2\Delta)\ \ \ \ U_2 = I_b \sin(2\chi+2\Delta)\\
Q_1&  =&  I_b \cos(2\chi-2\Delta)\ \ \ \ Q_2 = I_b \cos(2\chi+2\Delta)
\end{eqnarray}
The intensity factor $I_b$ will cancel out when calculating the net ${\rm
PA}=0.5\arctan(U/Q)$, so below we set $I_b=1$ to ignore it for simplicity. Thus we have:
\begin{eqnarray}
U&=&2\sin{2\chi}\cos{2\Delta}\\
Q&=&2\cos{2\chi}\cos{2\Delta}
\end{eqnarray}
so the factor $\cos{2\Delta}$ cancels out, and the result is:  
\begin{equation}
{\rm PA} = 0.5 \arctan{(U/Q)} = 0.5 \arctan{(\tan{2\chi})} = \chi. 
\label{resul1}
\end{equation}
As expected, a pair of two equally bright signals produces the net PA in the
middle between their individual polarization directions. The same result is
obtained in the opposite quadrant, ie.~for $135^\circ<\Delta<180^\circ$.
 
However, for the white bullets $\Delta$ exceeds $45^\circ$ so we can write $\Delta =
45^\circ+\delta$, with $0<\delta<90^\circ$ and, accordingly:
\begin{equation}
U_1 = I_w \sin(2\chi-90^\circ - 2\delta),\ \ \ U_2 =   I_w \sin(2\chi+90^\circ+ 2\delta)\\
\end{equation}
\begin{equation}
Q_1 = I_w \cos(2\chi-90^\circ - 2\delta),\ \ \ Q_2 =  I_w \cos(2\chi+90^\circ + 2\delta)\\
\end{equation}
Again we set $I_w=1$ for simplicity, and the usual properties of trigonometric functions 
 (eg.~$\sin(2\chi-90^\circ)=-\cos2\chi$, $\cos{(2\chi-90^\circ)}=\sin2\chi$, etc.) 
lead to 
\begin{eqnarray}
U&=&-2\sin{2\chi}\sin{2\delta}=2\sin{(2\chi\pm180^\circ)}\sin{2\delta}\\
Q&=&-2\cos{2\chi}\sin{2\delta}=2\cos{(2\chi\pm180^\circ)}\sin{2\delta}.
\end{eqnarray}
The value of $2\sin{2\delta}$ is positive for $\delta$ within the considered
range, 
so this factor may be neglected in the argument of atan2 function, hence: 
\begin{equation}
{\rm PA} = 0.5{\rm atan2}{\left(\sin{(2\chi\pm180^\circ)}, \cos{(2\chi\pm180^\circ)}\right)} = \chi\pm90^\circ
\label{resul2}
\end{equation}
to be compared with eq.~(\ref{resul1}). Indeed the orthogonal value has been
obtained. 
For a locally symmetric emission region all its emission points, when
considered in symmetric pairs, will then produce PA of either $\chi$ or $\chi\pm90^\circ$. 
The final net PA will depend on which contribution (specifically - its polarized part) is brighter, 
ie.~for which pair
of quadrants the polarized part of flux integrated within the entire area is larger. 
This implies that linear drop of
emissivity should cause essentially complete depolarization with a random PA
(noise), whereas less uniform decay of emissivity, or unfilled area around
the sightline (void of radiation) can produce higher average polarized fractions.

\section{Results}
\label{results}

\subsection{The conal case}
\label{ccase}

   \begin{figure}[ht!]
   \centering
   \includegraphics[width=0.9\hsize]{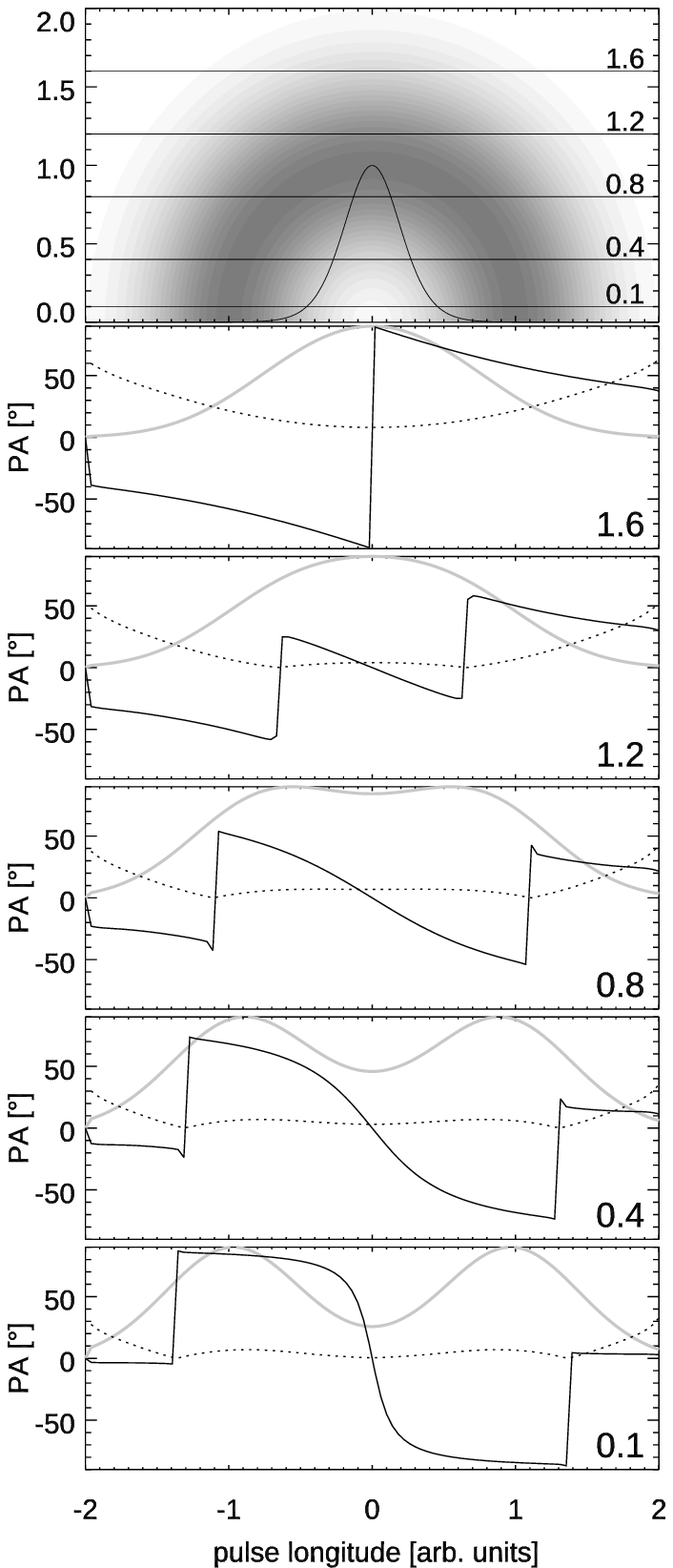}
      \caption{Polarization profiles for five sightline cuts through the hollow cone shown in grey in top
panel. The bell-shaped curve on top is the beamlet to be convolved with the cone's
emissivity. Bottom panels show the PA (solid line), $L/I$ (in percent, dotted) and the
intensity profile in arb.~units (thick grey). The sightline impact parameter
$y_{\rm los}$ 
is shown on the right and marked with horizontal lines on the beam map. Note
the RVM shaped PA curves and the OPM jumps.}
         \label{cthick}
   \end{figure}

   \begin{figure}[ht!]
   \centering
\includegraphics[width=0.9\hsize]{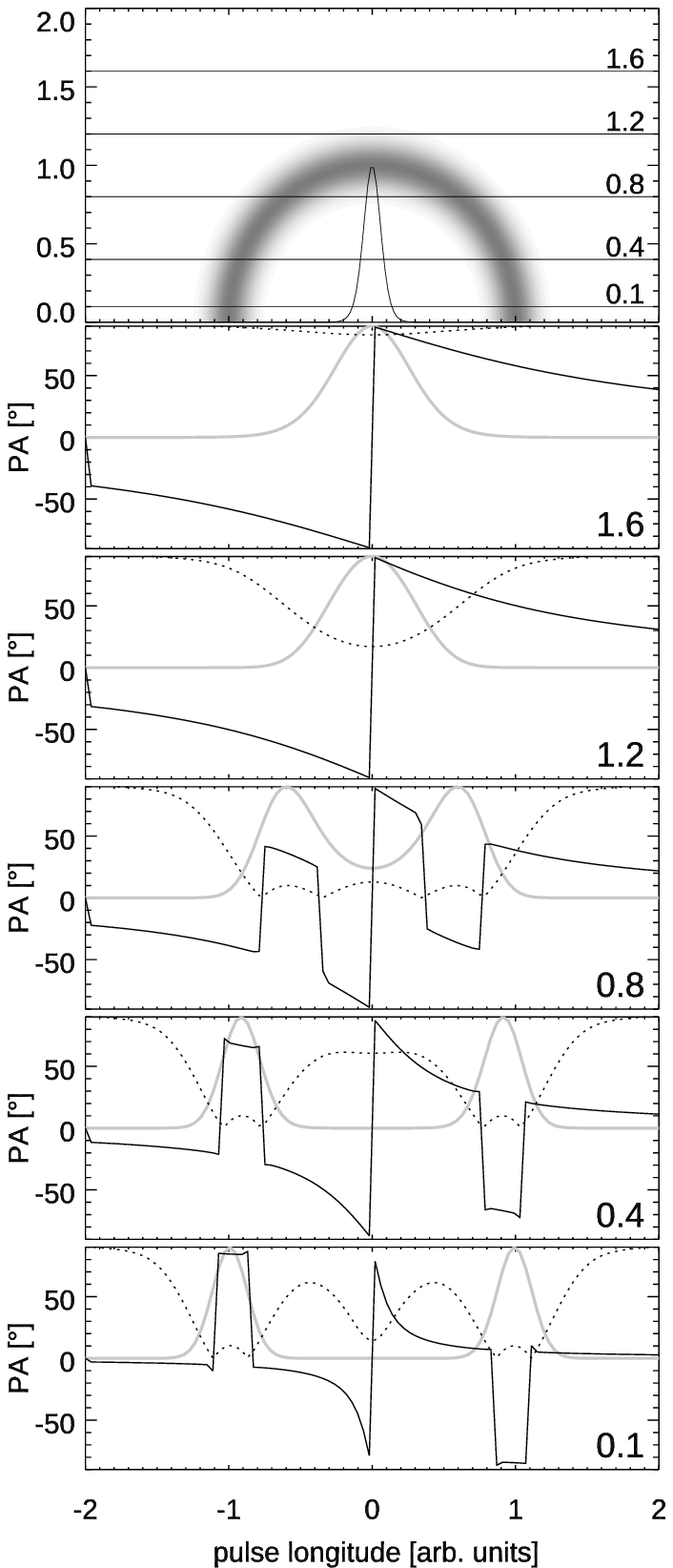}
      \caption{As in Fig.~\ref{cthick} but for a thinner cone with narrower
beamlet.}
         \label{cthin}
   \end{figure}

Figures \ref{cthick} and \ref{cthin} show polarization profiles 
for five sightline cuts through an arbitrary hollow cone, ie.~a cone that in
general is not centered at the dipole axis. Half of the modelled cone, assumed to
have radius $r_{\rm cn}=1$, is shown in the top
panel. The horizontal sightline paths have numbers equal to the path
distance $y_{\rm los}$ from the cone center (in 
units of $r_{\rm cn}$). Thus $y_{\rm los}$  corresponds to the impact angle in
the spherical
case, though now it is not measured from the dipole axis. 
The bell-shaped curve on top shows the 
beamlet size and shape - its axially symmetric version is anchored at each point of the conal emitter. 

In the case of Fig.~\ref{cthick}, we have used $\gamma=1.5$, and the cone wall
had a Gaussian shape with $\sigma_{\rm er}=0.4$. The first major feature of calculated
profiles is that the PA (solid line) follows the RVM curve with OMPs, although
the projected $B$ is not even present in the model (not directly, although one
may imagine that each point of the emitter is a sky-projected tip of a local $B$). 
The RVM with OPM jumps is the key observed property of radio pulsar
emission. 
Secondly, one can observe the specific relation between the form of the PA
curve and of the profile: with the steepening of central PA (and with the
increase of the overall PA interval) the profile becomes wider and double. 
This is a behaviour generally observed amoung radio pulsars and interpreted
as a more central cut through the cone. However, here this closer cut does not
automatically implies being closer to the dipole axis. 
Thirdly, for impact angles roughly between 0.8 and 1.3, the PA curve has the
triple form that is often observed in radio pulsars: the central PA track is
flanked by another `mode' shifted by $90^\circ$. The feature is for example
prominent in PSR J1110-5637, J1146-6039, J1210-5559, J1326-6700, 
J1456-6843, J1852-0635, J2139+2242 to name just a few objects from the
Thousand-Pulsar-Array programme
on MeerKAT (Johnston et al.~2020). \nct{jkk2020} 
For near-central cuts, the OPM jumps occur close to the profile edge, which
may be associated with observation of `edge depolarization' (Rankin \& Ramachandran
2003). \nct{rr03}
Fourthly, in terms of mode distance from the cone axis, 
the triple form (eg. for $y_{\rm los}=1.2$) implies a more peripheric
(outer) location of one mode (previously thought to be the refracted O mode) - 
however, the current calculation does not include any refraction.
Because of the wide cone wall and wide beamlet, the linearly polarized
fraction $L/I$ in Fig.~\ref{cthick} is very low, as happens to be observed in
several average pulsar profiles (eg.~in PSR J1456-6843, Dyks et al.~2021,
among many others). \nct{dwi21}

The results of the next Fig.~\ref{cthin} have been obtained for 
Gaussian cone walls that are much thinner ($\sigma_{\rm er}=0.1$) than the cone openning
angle) and for a narrower beamlet ($\gamma=5$). 
Therefore, the overall polarized fraction is here much larger.  
In addition to the above-described features, one can see the presence of another mode 
under weakly polarized peaks, with high polarized fraction in the bridge
between the peaks. The narrow longitude interval of another mode, also
happens to be observed in some objects, eg.~in the trailing component of PSR
B2020$+$28 -- see fig.~6 in DSF26 \nct{dsf2026} for the comparison to Arecibo data (Mitra et
al.~2015). \nct{mar2015}  
The linearly polarized fraction of the bridge emission is higher than in
peaks, which has also been noticed observationally (eg.~PSR B0521$+$21, Young \&
Rankin 2012). \nct{yr12} 
Within the bridge the modelled $L/I$ sometimes shows a drop of $L/I$ in the very center 
($y_{\rm los}=0.4$ and more pronouncedly for the near-central cut $y_{\rm
los}=0.1$). Similar feature is present in  PSR B0521$+$21. 

The modelled PA curves show that either OPM can dominate at the center of the profile, which 
depends both on the impact parameter (cf.~cases 1.6 and 1.2 in Fig.~\ref{cthick}) 
and the width of the cone wall and beamlet (cf.~cases 0.1 in Fig.~\ref{cthick} and
0.1 in Fig.~\ref{cthin}). This ambiguity of central mode
domination is in line with the observation of Johnston et al.~(2005) and
Rankin (2015) 
\nct{jhv05, r2015} 
that each mode can be aligned with the direction of pulsar proper motion.

\subsubsection{Interpretation of conal results}

\begin{figure}[ht!]
   \centering
   \includegraphics[width=\hsize]{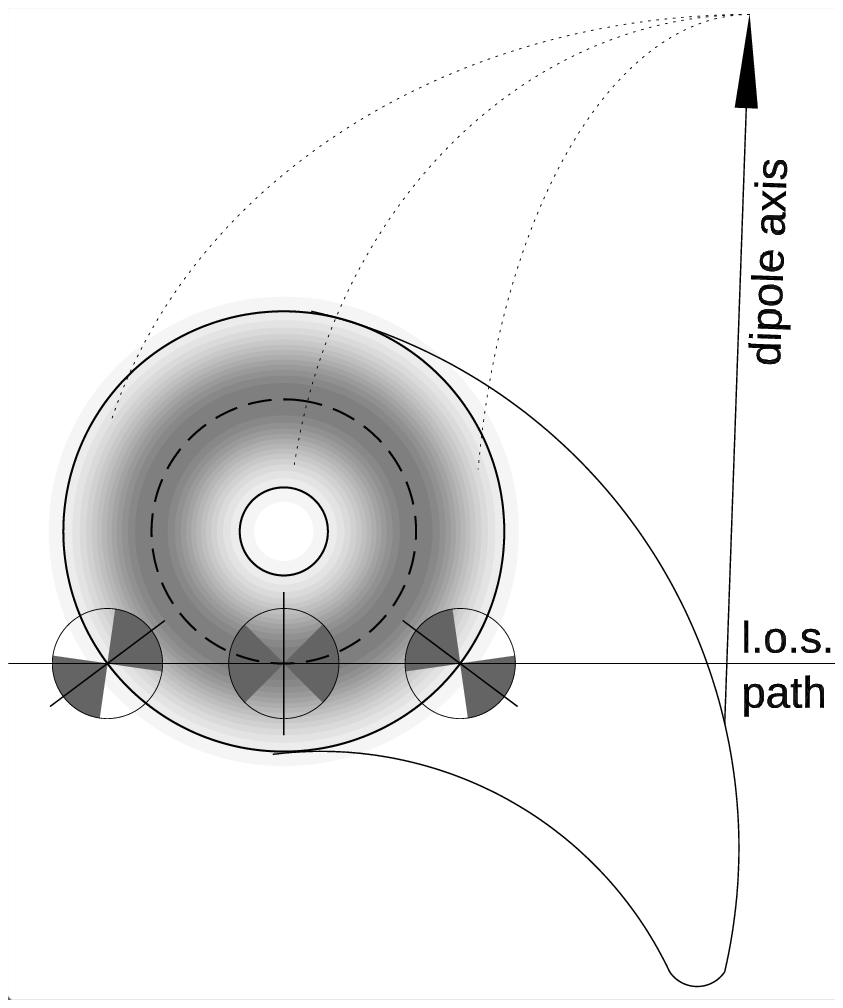}
      \caption{Example of an emission region with polarization of
Figs.~\ref{cthick} and \ref{cthin}. The maximum emissivity is at the dashed
line. The local probes of Fig.~\ref{lup} are shown at three positions on
sightline path. The probes reveal fast rotation of emissivity gradient that
mimicks the RVM 
(they always point to the beam's center) as well as the  
switch of polarization mode near the profile center (grey quadrants mark
the strongest mode). The beam is not centered at the dipole axis (dotted
lines mark fixed magnetic azimuth) so the standard RVM
would imply little or no change of PA.}
         \label{rob}
   \end{figure}

Fig.~\ref{rob} presents a tentative geometry of an emission region that
illustrates the origin of several results of Section \ref{ccase}. Most
emission comes from a circular crossection of a bent magnetic flux tube (a stream or a bunch of
magnetic field lines) with maximum emissivty at the dashed circle. 
Note that the circular emitter (the crossection) is not centered at the
dipole axis - the dotted arcs show the planes of a fixed magnetic azimuth that
in the traditional model would fix the PA. Here, since the stream is narrow,
there would be little or no RVM swing from projections of B-field planes
(the planes are nearly parallel within the emission region).

To assess the expected polarization it is helpful to use  
the geometry of Fig.~\ref{lup} - thus the circle of sightline vicinity with
the quadrants of orthogonality - as a local probe of polarization. 
Three such probes are drawn at the horizontal path of sightline in
Fig.~\ref{rob}. 
In the course of the sightline traverse through the beam, 
the probes keep facing the direction of the steepest gradient of emissivity.
So the Mode 1 axis of the probes rotates while always pointing towards
the cone center - thus mimicking the RVM (direction of Mode 1 is marked with the short
line section at each probe). For the left probe that is entering the beam,
the strongest emissivity mostly occupies the top right Mode 1 quadrant. For
the exiting probe on the right it is the top left quadrant of Mode 1 again. 
Hence Mode 1 should dominate the signal in the beam periphery. 
However, for the central probe the peak emissivity (dashed line) extends mostly in the
orthogonal quadrants of Mode 2 - hence there is another OPM in the beam centre. 
The Mode 2 is also located closer to the beam center (at a smaller
beam-centered colatitude) which is consistent with some traditional beam
mapping based on the carousel model (there, however, the colatitude was
assumed to be magnetic, ie.~centered at the dipole axis, Rankin \&
Ramachandran 2003; Deshpande \& Rankin 2001). \nct{rr03, dr2001} 
As mentioned above, the colatitude segregation in this model here is not caused by the refraction.

The use of local probes in Fig.~\ref{rob} thus helps to qualitatively understand most of
the polarization behaviour described above (RVM curve that is not determined
by the planes of magnetic azimuth, OPM jumps,
triple form of mode separation, smaller colatitude of the central mode). 
However, it must be emphasized that in general the quick assessment based on local probes may be
misleading, eg.~when the emission region is asymmetric or nonuniform on scales smaller
than the used probe size. Moreover, it must be remembered that it is
only the polarized part of the flux contributed by each pair of symmetric
points (Fig.~\ref{lup}) that counts when determining which mode dominates. 

\subsection{OPMs in single pulses}

Like in DSF26, the model presented here is based on incoherent superposition of polarized signal, 
which is correct at least for polarization of averaged profiles, because radiative contributions come
from different pulsar rotations (and it is the observers who do the
superposition with the help of Stokes parameters). 
However, the incoherent superposition was considered an issue in DSF26, 
who focused on a signal from passage of a single charge
(or bunch) that was moving along a bent $\vec B$-field line. 
Nevertheless, the elongated one-dimensional emitter of DSF26 can be amended by assuming that
the elongation is in the colatitude direction, ie.~occupying neighboring magnetic
flux tubes instead of following a single line or tube. This allows us to extend the
incoherent summation rule to the single pulse emission. 
Such elongated emitters can be active for a very short time and, in
different pulse periods, they can appear at  
different distances to the line of sight. Since the observed mode depends on
this distance (see fig.~5 in DSF26), either one or another orthogonal
PA value will be observed. Thus, as long as the temporary, elongated emitters are 
 assumed as the basic emitting entity, there is no problem for the nonbirefringent model to explain the
appearance of different orthogonal modes at a fixed pulse longitude in
different rotation periods. This fact has several applications to known
pulsar polarization properties, which 
will be
presented elsewhere.

\section{Conclusions}
\label{discu}

The nonbirefringent model naturally produces a number of basic polarization
properties observed in average profiles of radio pulsars. 
In particular - while not being based on planes of dipolar magnetic azimuth
- in one shot it implies RVM shaped PA curve with OPM jumps. The model also 
implies the often-observed triple form of mode separation, as well as is capable of 
producing different OPMs at a given pulse longitude in different pulsar
periods. This strongly
supports the idea that pulsar's OPMs have nonbirefringent nature, as
previously envisaged in DSF26. \nct{dsf2026}
The similarity of model results to several known observed
pulsar properties suggests
that the mechanism of nonbirefringent OPMs is not a subtle complication -
instead it is vital for the observed polarization. Altogether, the
presented results point to a picture in which 
the observed pulsar OPM's (orthogonal PA tracks) have nothing to do with the
two proper modes of wave oscillation - instead they result from a simple geometric effect operating on a
single proper mode. That is, it is the O-mode forming both tracks. Such a model of modes opens 
completely new ways of interpreting radio pulsar polarization.

\begin{acknowledgements}
We thank Aris Karastergiou and Michael Keith for convenient views of  
MeerKAT TPA pulsar data. 
This research was funded by National Science Centre, Poland, grant 
no.~2023/49/B/ST9/01783.  
For the purpose of Open Access, the author has applied a CC-BY public
copyright licence to any Author Accepted Manuscript version arising from
this submission.
\end{acknowledgements}

%
\bibliographystyle{aa} 
\bibliography{listofrefs2} 

\end{document}